\documentclass[12pt]{article} 
\usepackage{graphicx}
\usepackage{amsbsy}

\begin{document}

\centerline{\large \bf Charge qubit rotations in a double-dot nanostructure}

\vskip 4mm

\centerline{Leonid A. Openov and Alexander V. Tsukanov$^{\dag}$}

\vskip 2mm

\centerline{\it Moscow Engineering Physics Institute (State University),}
\centerline{\it Kashirskoe sh. 31, Moscow 115409, Russia}
\centerline{\it e-mail: opn@supercon.mephi.ru}

\vskip 2mm

\centerline{\it $^{\dag}$Institute of Physics and Technology RAS,}
\centerline{\it Nakhimovsky pr. 34, Moscow 117218, Russia}
\centerline{\it e-mail: tsukanov@ftian.oivta.ru}

\vskip 4mm

\begin{quotation}

Quantum operations with a charge solid-state qubit whose logical states are
formed by two spatially separated localized states of an electron in the
double-dot structure are studied theoretically. We show that it is possible
to perform various one-qubit rotations making use of the microwave
pulses tuned to the resonances between the localized states and the excited
state delocalized over the nanostructure. Explicit analytic expression for
the time-dependent electron state vector is derived, and the appropriate
pulse parameters are determined.

\end{quotation}

\vskip 4mm

PACS Numbers: 03.67.Lx, 73.21.La, 85.35.-p

\vskip 6mm

During past decade, quantum computing (QC) or, more generally, quantum
information processing, attracted much attention \cite{Nielsen}. The reasons
for that interest are (i) the existence of quantum algorithms  \cite{Barenco}
that could perform calculations exponentially faster as compared with
classical ones (QC software), and (ii) the rapid development of technology
and material science \cite{Dowling} that allowed to realize some prototypes
of QC devices (QC hardware). The key elements of the QC
hardware are the quantum bit or {\it qubit}, which is a generic two-level
system, and a register of such qubits. The register allows to store the
quantum information, which is processed by means of unitary transformations
({\it quantum gates}) through an external control.

Quite generally, any two-level system that has sufficiently long-lived
states and allows for efficient readout can be used for QC. A lot of
proposals for qubit realization have been made, see, e. g.,
Refs. \cite{Nielsen,Barenco,Loss,Kane,Valiev,Pashkin,Hollenberg}.
For practical
applications one has to look for a physical system that could serve as a base
for a {\it scalable} QC. It is commonly believed that the problem of
scalability can be effectively solved with the solid-state systems. In this
work, we focus on the nanostructure consisted of two quantum dots (QDs) and
containing an excess electron in the superposition of the orbital states
localized in different QDs \cite{Openov,Valiev}.
Those states form the logical
states of the charge qubit. We show that various one-qubit rotations,
including the phase gate, the NOT operation, and the Hadamard transformation
can be performed on such a qubit by means of one (if QDs are identical) or,
generally, two (if QDs differ from each other) microwave pulses tuned to the
resonances with one of the excited states delocalized over the nanostructure.

Let us consider the nanostructure composed of two QDs (L and R) and
containing an excess electron in the conduction band. Provided that
the distance between the QDs is sufficiently large, the wave functions
$\langle {\bf r}|L\rangle$ and $\langle {\bf r}|R\rangle$ of the lowest
size-quantized QD
states $|L\rangle$ and $|R\rangle$ with the energies $\varepsilon_L$ and
$\varepsilon_R$, respectively, are localized in the corresponding QDs, and
the overlap $\langle L|R\rangle$ is negligibly small. We assume that there is
at least one excited state $|TR\rangle$ in the nanostructure whose energy
$\varepsilon_{TR}$ lies just below
the top of the potential barrier separating the QDs (see Fig. 1),
so that the wave function $\langle {\bf r}|TR\rangle$ is delocalized
over both QDs \cite{Tsukanov}. Previously \cite{Openov,Tsukanov} we have
shown that in the case of the symmetric double-dot nanostructure,
$\varepsilon_L=\varepsilon_R$, the resonant electromagnetic pulse with the
frequency $\omega=\varepsilon_{TR}-\varepsilon_{L,R}$ (hereafter $\hbar=1$)
can be used to achieve the complete population transfer between the states
$|L\rangle$ and $|R\rangle$. In such a process, an excited state $|TR\rangle$
plays the role of the "transport" state: it assists the electron transfer
between the QDs but remains unpopulated after the pulse is off. Here we show
that in the double-dot nanostructure various
types of the temporal evolution of a {\it superpositional} state
$\alpha|L\rangle+\beta|R\rangle$ can be realized in both
$\varepsilon_L=\varepsilon_R$ and $\varepsilon_L\neq\varepsilon_R$ cases
through the proper choice of the pulse parameters.

We subject the nanostructure to the external electromagnetic pulse of the
form
\begin{equation}
{\bf E}(t)=\biggl[{\bf E}_0\cos(\omega_0 t+\varphi_0)+
{\bf E}_1\cos(\omega_1 t+\varphi_1)\biggr]\cdot
\biggl[\theta(t)-\theta(t-T)\biggr]~,
\label{E(t)}
\end{equation}
where $T$ is the pulse duration, i. e., in fact, the operation time. We
assume that the frequency $\omega_{0,1}$ is close to the resonant frequency
$\omega_{L,R}=\varepsilon_{TR}-\varepsilon_{L,R}$ for the electron transition
$|L,R\rangle\rightleftharpoons |TR\rangle$, so that
$|\delta_{0,1}|<<\omega_{0,1}$, where
$\delta_{0,1}=\omega_{L,R}-\omega_{0,1}$ is the corresponding detuning
(see Fig. 1). In the resonant approximation (see Refs. \cite{Openov} and
\cite{Tsukanov} for details), only the states $|L\rangle$, $|R\rangle$, and
$|TR\rangle$ participate the electron evolution, so that the problem reduces
to a three-level model, and the state vector
$|\Psi(t)\rangle$ can be written as
\begin{equation}
|\Psi(t)\rangle = \sum_{k=L,R,TR}C_k(t)e^{-i\varepsilon_k t}|k\rangle~,
\label{Psi(t)}
\end{equation}
and the Hamiltonian reads
\begin{equation}
H(t)=\sum_{k=L,R,TR}\varepsilon _k a_k^+ a_k^{}+
\frac{1}{2}\biggl(\lambda_L a_{TR}^+ a_L^{} e^{-i\omega_0 t}
+\lambda_R a_{TR}^+ a_R^{}e^{-i\omega_1 t}+h.c.\biggr)~,
\label{H}
\end{equation}
where $a_k^+$ ($a_k^{}$) is the operator of creation (annihilation) of an
electron in the state $|k\rangle$,
$\lambda_{L,R}={\bf E}_{0,1}{\bf d}_{L,R}e^{-i\varphi_{0,1}}$, and
${\bf d}_{L,R}=-e\langle TR|{\bf r}|L,R\rangle$ is the dipole matrix element
for the transitions $|L,R\rangle\rightleftharpoons |TR\rangle$.

To solve the non-stationary Schr\"{o}dinger equation
\begin{equation}
i\frac{\partial |\Psi(t)\rangle}{\partial t}=H(t)|\Psi(t)\rangle
\label{Non-stat}
\end{equation}
with the Hamiltonian (\ref{H}) and the initial conditions
\begin{equation}
|\Psi(0)\rangle = \alpha|L\rangle + \beta|R\rangle~,
\label{Initial}
\end{equation}
we make use of the unitary transformation
\begin{equation}
|\Psi(t)\rangle = U(t)|\tilde{\Psi}(t)\rangle
\label{PsiU}
\end{equation}
with
\begin{equation}
U(t)=\exp\biggl(i\omega_0 ta_L^+a_L^{}+i\omega _1 ta_R^+a_R{}\biggr)~.
\label{U}
\end{equation}
Substituting Eq. (\ref{PsiU}) into Eq. (\ref{Non-stat}), we obtain the
Schr\"{o}dinger equation for $|\tilde{\Psi}(t)\rangle$:
\begin{equation}
i\frac{\partial |\tilde{\Psi}(t)\rangle}{\partial t}=
\tilde{H}|\tilde{\Psi}(t)\rangle~,
\label{Non-stat2}
\end{equation}
with the Hamiltonian $\tilde{H}$ in the basis
$\{|L\rangle,|R\rangle,|TR\rangle\}$ being
\begin{equation}
\tilde{H}=U^+(t)H(t)U(t)-iU^+(t)\frac{\partial U(t)}{\partial t}=
\left( {\begin{array}{*{20}c}
{\varepsilon_{TR}-\delta_0 } & 0 & {\frac{{\lambda _L^* }}{2}}  \\
0 & {\varepsilon_{TR}-\delta_1} & {\frac{{\lambda _R^* }}{2}}  \\
{\frac{{\lambda _L^{} }}{2}} & {\frac{{\lambda _R^{} }}{2}} &
{\varepsilon_{TR}}  \\
\end{array}} \right)~.
\label{Htilde}
\end{equation}
Since the Hamiltonian $\tilde{H}$ is time-independent, the general solution
of Eq. (\ref{Non-stat2}) for $0\le t\le T$ is
\begin{equation}
\tilde{\Psi}(t)=
\sum_{k=1}^{3} A_k e^{-i\tilde{\varepsilon}_kt}|\tilde{k}\rangle~,
\label{Psitilde}
\end{equation}
where $|\tilde{k}\rangle$ and $\tilde{\varepsilon}_k$ are, respectively, the
eigenstates and eigenenergies of the stationary Schr\"odinger equation
\begin{equation}
\tilde{H}|\tilde{k}\rangle=\tilde{\varepsilon}_k|\tilde{k}\rangle~.
\label{Stat}
\end{equation}
They can be found from the cubic equation for eigenvalues of the 3x3 matrix
(\ref{Htilde}).

The general features of the qubit evolution in a three-level model
have been discussed in Ref. \cite{Amin} by the example of the Josephson phase
qubit. Here we consider an important particular case of exact resonances,
$\delta_0=\delta_1=0$, that allows for a simple analytical solution.
In this case one has
\begin{equation}
\tilde{\varepsilon}_1=\varepsilon_{TR}-2\Omega,~
\tilde{\varepsilon}_2=\varepsilon_{TR},~
\tilde{\varepsilon}_1=\varepsilon_{TR}+2\Omega~,
\label{Etilde}
\end{equation}
\begin{eqnarray}
&& |\tilde{1}\rangle=
\frac{1}{\sqrt{2}}\biggl[-\frac{\lambda_L^*}{4\Omega}|L\rangle
-\frac{\lambda_R^*}{4\Omega}|R\rangle+|TR\rangle\biggr]~,
\nonumber
\\ && |\tilde{2}\rangle=\frac{1}{4\Omega}\biggl[\lambda_R|L\rangle
-\lambda_L|R\rangle\biggr]~,
\nonumber
\\ && |\tilde{3}\rangle=
\frac{1}{\sqrt{2}}\biggl[\frac{\lambda_L^*}{4\Omega}|L\rangle
+\frac{\lambda_R^*}{4\Omega}|R\rangle+|TR\rangle\biggr]~,
\label{ktilde}
\end{eqnarray}
where
\begin{equation}
\Omega=\frac{\sqrt{|\lambda_L|^2+|\lambda_R|^2}}{4}
\label{Omega}
\end{equation}
is the Rabi frequency for the system under consideration.

Taking into account that $|\tilde{\Psi}(0)\rangle=|\Psi(0)\rangle$,
it is straightforward to find the coefficients $A_k$ in Eq. (\ref{Psitilde})
from the initial condition (\ref{Initial}). Then we have from
Eqs. (\ref{PsiU}) and (\ref{U}) the explicit expression for the state vector
$|\Psi(t)\rangle$:
\begin{eqnarray}
&& |\Psi(t)\rangle=e^{-i\varepsilon_L t}
\biggl[\alpha-\frac{\lambda_L^*(\alpha\lambda_L+\beta\lambda_R)}{8\Omega^2}
\sin^2(\Omega t)\biggr]|L\rangle
\nonumber
\\ &&
+e^{-i\varepsilon_R t}
\biggl[\beta-\frac{\lambda_R^*(\alpha\lambda_L+\beta\lambda_R)}{8\Omega^2}
\sin^2(\Omega t)\biggr]|R\rangle
\nonumber
\\ &&
-ie^{-i\varepsilon_{TR} t}
\frac{\alpha\lambda_L+\beta\lambda_R}{4\Omega}\sin(2\Omega t)|TR\rangle~.
\label{Psi(t)2}
\end{eqnarray}
One can see from Eq. (\ref{Psi(t)2}) that at operation times
$T_n=\pi n/2\Omega$ ($n$ = 1, 2, ...) the state vector is completely
localized in the logical qubit subspace $\{|L\rangle,|R\rangle\}$. In this
sense, the qubit evolution appears to be stroboscopic with respect to the
logical subspace.

Now let us demonstrate how various qubit rotations can be performed through
the proper choice of the pulse parameters.
At $T=\pi k/\Omega$ ($k$ = 1, 2, ...) the relative phase shift operation is
realized provided $\varepsilon_L\neq\varepsilon_R$,
\begin{equation}
|\Psi(T)\rangle = e^{-i\varepsilon_L T}\biggl[\alpha|L\rangle+
\beta e^{-i(\varepsilon_R-\varepsilon_L)T}|R\rangle \biggr] ~.
\label{Phase}
\end{equation}
At $T=\pi (2k-1)/2\Omega$ ($k$ = 1, 2, ...) one has
\begin{eqnarray}
&& |\Psi(T)\rangle = e^{-i\varepsilon_L T}\biggl[
\alpha\frac{|\lambda_R|^2-|\lambda_L|^2}{|\lambda_R|^2+|\lambda_L|^2}
-\beta\frac{2\lambda_L^*\lambda_R^{}}{|\lambda_R|^2+|\lambda_L|^2}
\biggr]|L\rangle
\nonumber
\\ &&
+e^{-i\varepsilon_R T}\biggl[
\beta\frac{|\lambda_L|^2-|\lambda_R|^2}{|\lambda_R|^2+|\lambda_L|^2}
-\alpha\frac{2\lambda_L^{}\lambda_R^*}{|\lambda_R|^2+|\lambda_L|^2}
\biggr]|R\rangle~,
\label{NOT_HAD}
\end{eqnarray}
so that the NOT gate,
\begin{equation}
|\Psi(T)\rangle = \pm e^{-i(\varepsilon_L+\varepsilon_R)T/2}
\biggl[\beta|L\rangle+\alpha|R\rangle\biggr]~,
\label{NOT}
\end{equation}
is implemented if $|\lambda_L|=|\lambda_R|$ and
$\varphi_1-\varphi_0=\pi n + (\varepsilon_R-\varepsilon_L)T/2$ ($n$ is an
integer). Note that the NOT gate can be realized in both asymmetric
$\varepsilon_L\neq\varepsilon_R$ and symmetric $\varepsilon_L=\varepsilon_R$
nanostructures. In the latter case there is no need in the second component
of ${\bf E}(t)$, see Eq. (\ref{E(t)}). Next, it follows from
Eq. (\ref{NOT_HAD}) that the Hadamard gate,
\begin{equation}
|\Psi(T)\rangle = \pm e^{-i\varepsilon_L T}
\biggl[\frac{\alpha+\beta}{\sqrt{2}}|L\rangle +
\frac{\alpha-\beta}{\sqrt{2}}|R\rangle\biggr]~,
\label{HAD}
\end{equation}
is realized if $(\varepsilon_R-\varepsilon_L)T=2\pi n$ ($n\neq 0$ is an
integer), $\varphi_1-\varphi_0=\pi m$ ($m$ is an integer), and
$\lambda_L/\lambda_R=1\pm\sqrt{2}$.

We note that the values of
$\varepsilon_L$ and $\varepsilon_R$ can be varied independently by applying
the voltage biases to the surface gates, while the values of $\lambda_L$ and
$\lambda_R$ can be adjusted through the appropriate changes in the electric
field amplitudes. So, various rotations of a charge qubit in the QD
nanostructure can be implemented through appropriate choice of the pulse
parameters (frequencies, phases, intensity, duration) and proper
nanostructure engineering. The unavoidable small deviations of the pulse
frequencies from the ideal resonance conditions and the departures of
$\lambda_{L,R}$ from their optimal values can be accounted for in a way
similar to that used in Ref. \cite{Tsukanov} for the resonant electron
transfer between the QDs.

The characteristic operation times are
$T\sim 1/(\varepsilon_R-\varepsilon_L)\sim 1$  ps for
$\varepsilon_R-\varepsilon_L\sim 1$ meV. Since
$T \sim 1/\Omega \sim 1/|\lambda_{L,R}| \sim 1/eaE_0$, where $a \sim 10$ nm
is the QD size, such values of $T$ correspond to the electric field
strength $E_0 \sim 10^3$ V/cm. The short operation times allow to minimize
the unwanted decoherence effects. Indeed, the lowest bounds for the
decoherence times due to the Nyquist-Johnson noise from the gates and 1/f
noise from the background charge fluctuations are \cite{Hollenberg}
$\tau\sim 1$ $\mu$s and $\tau\sim 1$ ns, respectively, so that the
corresponding error rates \cite{Fedichkin}
$D(T)=1-\exp(-T/\tau)$ do not exceed $10^{-3}$. As for the phonon-induced
decoherence, in the case that the absolute value of the difference
$\varepsilon_R-\varepsilon_L$ greatly exceeds the energy of electron
tunneling between the QDs (i. e., in the case of well separated QDs),
the main contribution to decoherence come from dephasing processes, and the
error rate in common semiconductors
is $D(t)=10^{-4}\div10^{-3}$ \cite{Fedichkin}. Such values of
$D(t)$ are close to the fault-tolerant threshold for quantum computation
\cite{DiVincenzo}. The detailed analysis of the decoherence effects is,
however, beyond the scope of this paper.

To summarize, we have shown that making use of the microwave pulses allows
one to implement various operations on a charge qubit encoded in two
spatially separated states of an electron in the double-dot nanostructure.
The extremely short operation times $\sim 1$ ps make it possible to minimize
the decoherence effects below the level sufficient for at least the
proof-of-principle experiments and demonstration of the feasibility of the
scheme discussed. Although we restricted ourselves to the rectangular pulses,
our treatment can be generalized to other pulse shapes \cite{Paspalakis}. The
results obtained can be applied also to the Josephson three-level gates
\cite{Amin,Yang,Kis}. Finally, it does not seem unrealistic to organize the
coupling of QDs-based charge qubits for conditional quantum operations.

Discussions with L. Fedichkin are gratefully acknowledged.

\newpage

\includegraphics[width=\hsize]{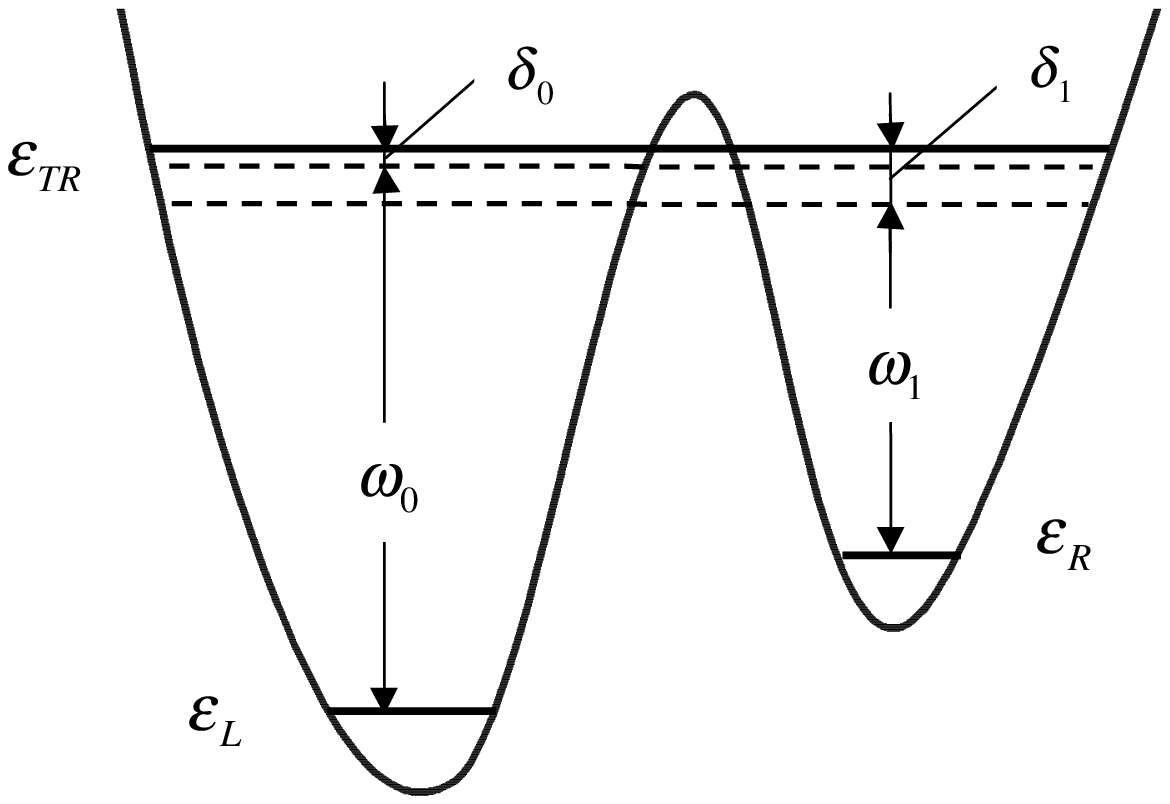}

\vskip 6mm

Fig. 1. Schematics of energy levels and resonant frequencies for the
double-dot nanostructure driven by the microwave pulse, see text for details.

\end{document}